\documentclass[acmsmall,screen]{acmart}

\usepackage{url}  
\usepackage{graphicx}  
\usepackage{wrapfig}  
\usepackage{multicol}
\usepackage[linesnumbered,ruled,vlined]{algorithm2e}
\usepackage{algorithmic}
\usepackage{graphicx}
\usepackage{textcomp}
\usepackage{xcolor}
\def\BibTeX{{\rm B\kern-.05em{\sc i\kern-.025em b}\kern-.08em
    T\kern-.1667em\lower.7ex\hbox{E}\kern-.125emX}}

\newcommand{\eat}[1]{}
\newcommand{\system}{ASTRA}

\usepackage{url}
\usepackage{wrapfig}
\usepackage{amsmath}
\usepackage{graphics}

\usepackage{graphicx}
\usepackage{caption}
\usepackage{subcaption}
\usepackage{array}
\usepackage{enumitem}
\newcolumntype{M}[1]{>{\centering\arraybackslash}m{#1}}
\usepackage{multirow}

\usepackage{url}
\usepackage{listings,color,colortbl}

\definecolor{light-gray}{gray}{0.80}
\def\denseitems{
  \itemsep1pt plus1pt minus1pt
  \parsep0pt plus0pt
  \parskip0pt\topsep0pt}
  
\usepackage{flushend}
\usepackage{balance}
\DeclareMathOperator*{\argmax}{argmax} 

\newcommand{\id}[1]{\mbox{\textit #1\/}}
\setcopyright{none}
\renewcommand\footnotetextcopyrightpermission[1]{}
\begin{document}

\title{Utilizing API Response for Test Refinement}
\author{Devika Sondhi}
\affiliation{%
  \institution{IBM Research}
  \country{India}
}
\email{devika.sondhi@ibm.com}

\author{Ananya Sharma}
\affiliation{%
  \institution{IIIT Delhi}
  \country{India}
}
\email{ananya20359@iiitd.ac.in}

\author{Diptikalyan Saha}
\affiliation{%
  \institution{IBM Research}
  \country{India}
}
\email{diptsaha@in.ibm.com}

\begin{abstract}
Most of the web services are offered in the form of RESTful APIs. This has led to an active research interest in API testing to ensure the reliability of these services. While most of the testing techniques proposed in the past rely on the API specification to generate the test cases, a major limitation of such an approach is that in the case of an incomplete or inconsistent specification, the test cases may not be realistic in nature and would result in a lot of 4xx response due to invalid input. This is indicative of poor test quality. Learning-based approaches may learn about valid inputs but often require a large number of request-response pairs to learn the constraints, making it infeasible to be readily used in the industry. To address this limitation, this paper proposes a dynamic test refinement approach that leverages the response message. The response is used to infer the point in the API testing flow where a test scenario fix is required. Using an intelligent agent, the approach adds constraints to the API specification that are further used to generate a test scenario accounting for the learned constraint from the response. Following a greedy approach, the iterative learning and refinement of test scenarios are obtained from the API testing system. The proposed approach led to a decrease in the number of 4xx responses, taking a step closer to generating more realistic test cases with high coverage that would aid in functional testing. A high coverage was obtained from a lesser number of API requests, as compared with the state-of-the-art search-based API Testing tools.


\end{abstract}

\maketitle


\section{Introduction}
\label{sec:intro}


OpenAPI \cite{swagger} specification files are often inconsistent with the API's behavior; this may happen due to incorrectly or incompletely written specifications, or these specifications may become inconsistent in the API evolution cycle. Further, certain behavioral constraints in the API may not be structurally supported in the API specification format.  For example, the specification cannot capture the associativity between operation parameters. However, knowing such missing constraints would certainly help automated API testers create valid or functional test cases. 

API testing techniques (e.g.,~\cite{arcuri2020automated, russo2022assessing, viglianisi2020resestgen, atlidakis2019restler, laranjeiro2021black, martin2021restest, corradini2022automated, corradini2021empirical, tsai2021rest, liu2022morest, wu2022combinatorial, segura2017metamorphic, stallenberg2021improving, karlsson2020quickrest, ed2018automatic}) can be broadly classified into white-box and black-box, based on the access to the API's artifacts. Due to its ease of use and setup and general applicability, black-box testing~\cite{atlidakis2019restler, liu2022morest} has been increasingly popular. The past techniques on black-box API testing have leveraged the input characteristics, such as parameter types, schema structure, and constraints such as minimum, maximum, enum, etc. in the OpenAPI specification to generate the test cases. They have also performed some inferences on API specification to derive important relationships such as producer-consumer relationships~\cite{atlidakis2019restler}. Recent works have also explored utilizing the description tags in OpenAPI specification to apply natural language processing to infer constraints to generate the test cases~\cite{kim2023nlp2rest}. 

However, the limitation of the above approaches is that they heavily rely on the specification to be correct or consistent with the API behavior, resulting in invalid test cases (4xx response code). Section \ref{sec:overview} shows one such test case. In such a case, it becomes important to take feedback from the API's behavior to refine the test case generation. Existing black-box techniques such as MOREST \cite{liu2022morest} use the response code to dynamically guide the call sequence generation. However, they still produce invalid test cases. Another adaptive API testing approach applies reinforcement learning to enhance the exploration strategy in a tool called ARAT-RL ~\cite{kim2023adaptive}. However, these search-based approaches require a large amount of data to learn enough information. This is evident from the large number of API requests that these tools make to be able to attain good coverage. While these techniques may be effective in obtaining good coverage, in the industry scenario, making a large number of API requests incurs a cost that may not be practical \cite{gamez2019role, upadhyaya2016price}. For such tools to be widely used in the practical scenario, the key is to make minimal API requests, while maximizing the testing effectiveness. This effectiveness may be measured through metrics such as operation coverage, defects revealed, and line coverage.

To address this problem, we propose a feedback-driven testing tool, {\system} (\textbf{A}PI'\textbf{s} \textbf{T}est \textbf{R}efinement \textbf{A}utomation), which applies a greedy approach to minimize the API requests while focusing on learning from the failures observed on these requests. An underlying intelligent agent in {\system}, generates test cases intending to create valid (non-4XX) test cases and leverages the natural language response message, along with the response code, for the test refinement. Functional test case constitutes generating a valid sequence of API operations, a valid selection of parameters for each operation, and a valid data value for each parameter. The generation of test immensely benefits from the different types of constraints inferred from response messages. 

The main challenge is to understand the response messages and infer constraints in the presence of natural language ambiguities and contextual omissions. The next challenge is to effectively use such constraints along with the existing knowledge derived from specification to create valid sequences, parameters, and data scenarios. The next section discusses our approach and various constraints identified through the initial study and how {\system}  handles each. We use large language models to infer constraints from the response messages and express them in the extended specification model capable of representing all such constraints. 

This work makes the following contributions:
\begin{itemize}
    \item To the best of our knowledge this is the first automated black-box API testing system that automatically analyzes the response messages for the generation of valid (that yields non-4xx response code) test cases. 
    \item We present an LLM-based methodology for extracting various types of constraints from response messages followed by an LLM-based producer-consumer relationship inference and data generation. We believe LLM is well suited for these tasks. 
    \item We present a feedback-driven algorithm that uses response messages and status to generate test cases, while optimising on the API requests made by following a greedy approach.
    \item By using much lesser responses, {\system} shows a higher percentage of valid test cases, compared to the existing techniques, in the majority of the benchmarks.  
\end{itemize}


\section{Approach}
\label{sec:approach}
\begin{figure}[t]
    \centering
    \includegraphics[width=5in]{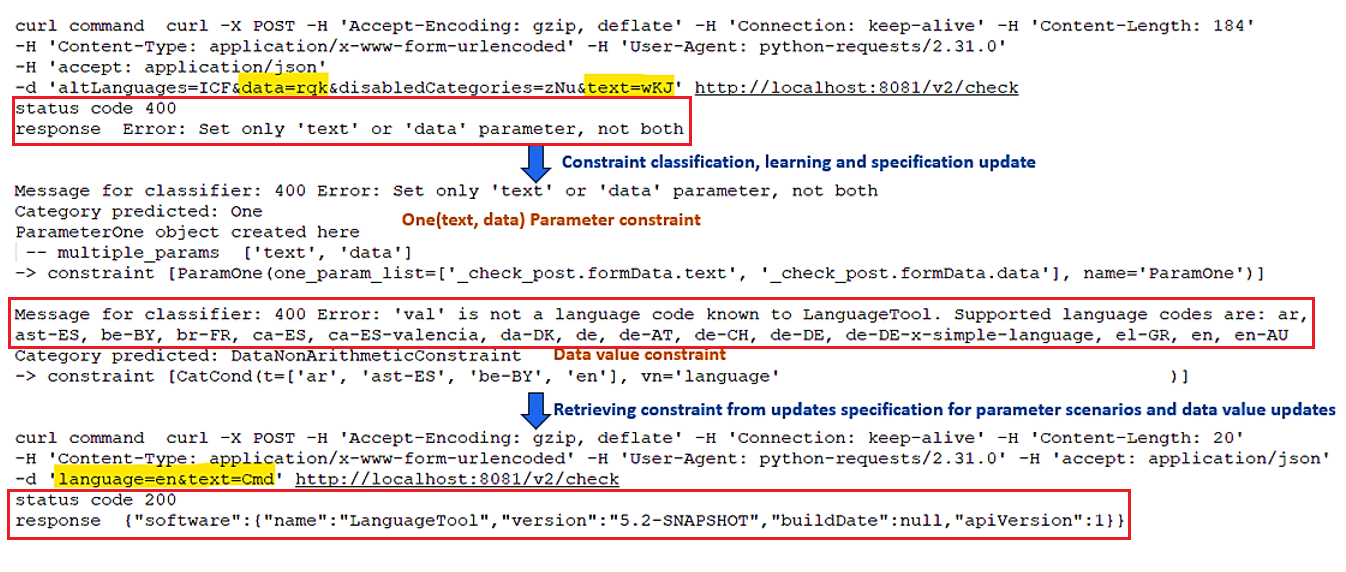}
    \vspace{-4mm}
    \caption{Sample Response 1}
    \label{fig:example}

\end{figure}

\eat{
\begin{figure}[t]
    \centering
    \includegraphics[width=2.5in]{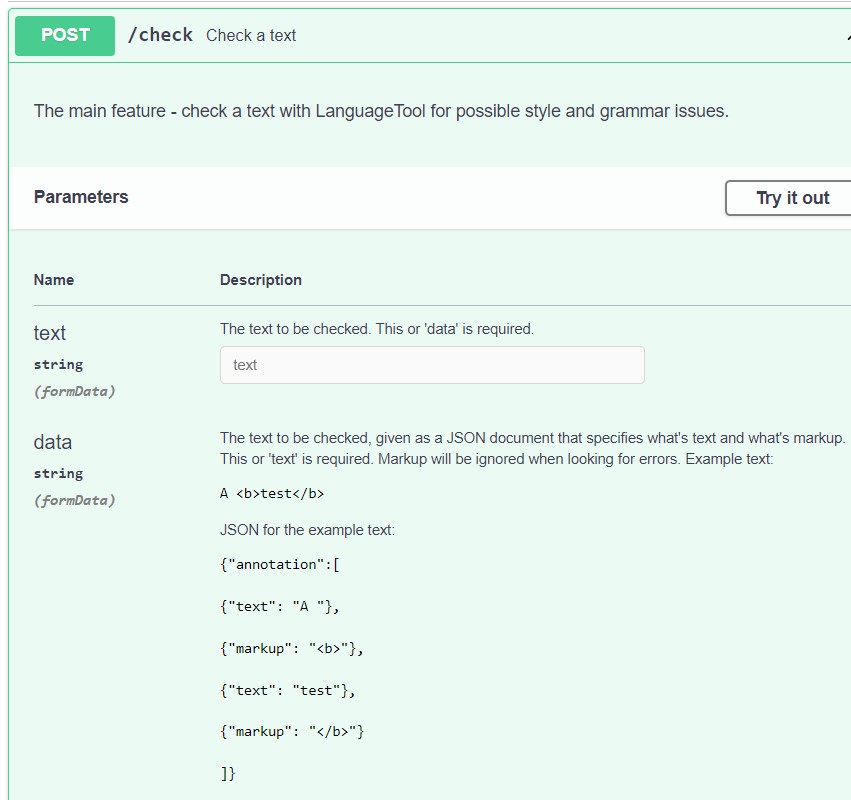}
    \caption{Snippet from the Language-tool API specification}
    \label{fig:spec}
\end{figure}
}
\eat{
\begin{figure}[t]
    \centering
    \includegraphics[width=3in]{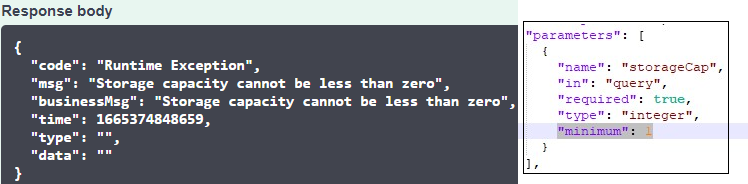}
    \caption{Sample Response 2}
    \label{fig:exampleIndirect}
\end{figure}

\begin{figure}[t]
    \centering
    \includegraphics[width=2in]{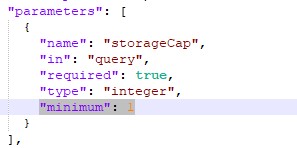}
    \caption{Snippet from the API specification}
    \label{fig:specIndirect}
\end{figure}
}


\subsection{Overview and Examples}\label{sec:overview}
\subsubsection{Resolving Parameter Constraints}
Fig. \ref{fig:example} shows two types of error response messages received from the two subsequent requests made to the Language-tool API\cite{languagetool}. The first response message indicates that for given optional parameters, exactly one of the two should be given a value. The request includes values for both the parameters, due to which the error 400 was thrown with a message. Note that the OpenAPI specification for this API contains no formal way to specify this inter-parameter constraint; its description states that either of them should be specified but nothing about whether specifying both is valid. So the first challenge is in recognizing the type of constraint from the natural language response. Another challenge is in identifying the entities to which the constraint applies. Among the list of parameters, the challenge is to identify that \texttt{text} and \texttt{data} are the parameters on which the identified constraint applies. At times, the parameter name may not directly appear in the response message. For instance, the response \textit{`Storage capacity cannot be less than zero'} specifies a minimum value constraint on the parameter \texttt{storageCap}. Hence, the second challenge is to automate parameter recognition, given there could be a long list of parameters that the operation takes as input. 

{\system} handles a response error by breaking the test refinement task into these subtasks- error category identification, target identification and actions to derive the constraint class. {\system} identifies the category of constraints from among a list of categories (Table \ref{tbl:classes}) identified through a preliminary study. This is done by an underlying LLM that learns the definition of various categories and picks the appropriate one for the given error message. The identification of category is essential for {\system} to determine at what level the test needs to be fixed. The levels could vary from sequence (operation), parameter or data. In the context of the example, one infers that the potential fix is in the parameter scenario of the operation and the identified category is \textit{One(p1, p2...)}. However, {\system} still needs to identify the arguments (p1, p2) for the constraint. {\system} then takes an appropriate action for the identified category. The action module is handled by an intelligent agent that allows identifying the target entities and derives a constraint. This results in inferring \texttt{text} and \texttt{data} as the target parameters.  So in this case, an instance of condition class taking the parameters is created as \textit{One(text, data)} and added to the specification model. These constraints are later consumed to generate the test scenarios containing the compliant parameter scenario in the request.

\subsubsection{Resolving Data Constraints}
Once the updated test case derived from the parameter constraint learnt from the previous error is executed, the Language API throws another error in the next run, as highlighted in the second message in Fig. \ref{fig:example}. The response message indicates that a data value passed in the request is not a supported one. {\system} identifies the category as a \textit{DataNonArithmeticConstraint(target, value, property)}. The agent of {\system} populates the constraint to identify that the target parameter is \texttt{language} with possible data values `ar', `en' etc. as Categorical values. The derived constraint is added to the specification model against the identified parameter. As can be seen at bottom in Fig. \ref{fig:example}, using the supported value `en' for \texttt{language} in the subsequent request results in a 200 response.
\begin{figure}[t]
    \centering
    \includegraphics[width=5in]{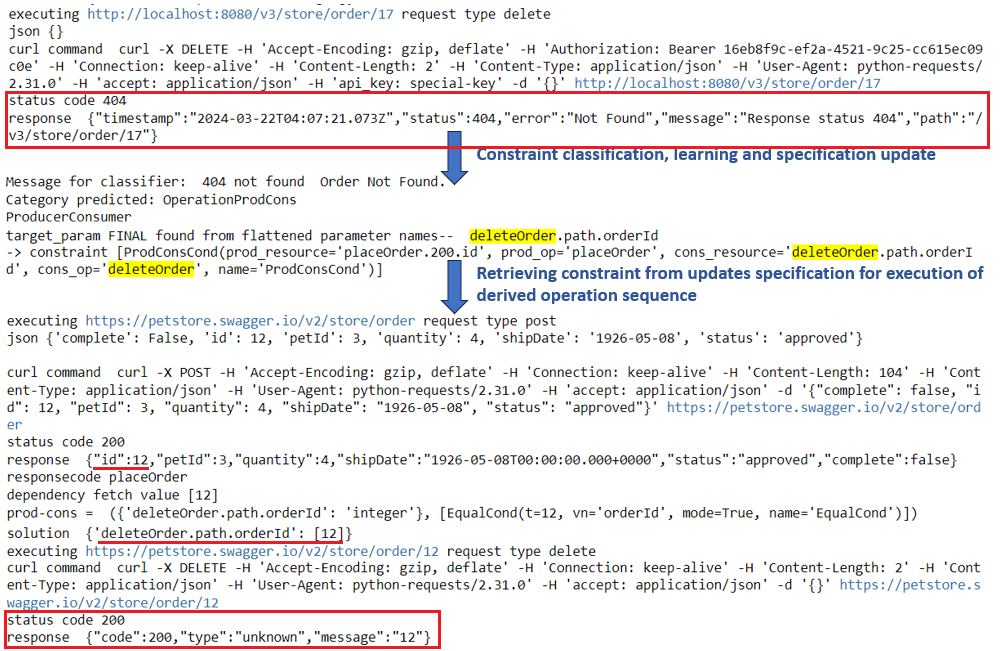}
    \caption{Sample Response 2}
    \label{fig:example2_op}
\end{figure}
\subsubsection{Resolving Operation Constraints}
Consider another example of an invalid request from the Petstore API\cite{petstore} in Fig. \ref{fig:example2_op}. The failed request returning a response 404 in the highlighted block is accompanied by a message `Order Not Found', indicating a missing \texttt{Order} resource. {\system} categorises this as a \textit{ProducerConsumer} constraint, given a pre-requisite operation was needed to be called to create that \texttt{Order} resource. The intelligent agent identifies the consumer parameter, the producer operation and the producer parameter to derive the constraint: \textit{ProducerConsumer(prodOp=placeOrder, prodParam=placeOrder.200.id, consOp=deleteOrder, consParam=deleteorder.path.orderId)} and adds it to the specification model. On the subsequent run, {\system} derives a sequence of operations \texttt{[placeOrder, deleteOrder]} from the constraint and injects the \texttt{id} produced by \texttt{placeOrder} to \texttt{deleteOrder}. This results in a successful 200 response, as can be seen in the figure.

To summarize, \system's  approach broadly comprises the following steps:
\begin{enumerate}
    \item Categorizing the logged errors into one of the identified constraint types. In total, we have identified 14 categories in which response messages can be classified. 10 of those categories yield constraints that can be used to refine the test cases.  Broadly, the constraints    
    may be on the operation, indicating the requirement of a pre-requisite operation; on the parameter level- indicating constraints on parameter occurrence, and on the data level- indicating the parameter input values that should be passed. Such constraints essentially impact three components of a test case - 1) sequence 2) parameter, and 3) data. 
    \item Identifying various types of constraints and updating the specification model with the inferred constraints. 
    \item Consuming the constraints from the updated specification model to generate sequence scenarios, parameter scenarios, and data scenarios to generate test cases.
\end{enumerate}

\subsection{Specification Model}
Before explaining the algorithm, we explain the Specification model, which is the data structure of the algorithm. The initial state of the model is obtained by parsing the OpenAPI specification. The specification model extends the OpenAPI specification information by including various constraint types (discussed later) which cannot be formally expressed in the OpenAPI specification. For example, exactly one of the two optional parameters should be passed in the operation request or the value of one parameter should be unique. 

The model is denoted as $Spec := \langle O, S\rangle$, where $O$ is the set of operations and $S$ is the set of schemas. Each operation $o \in O$ is a tuple $\langle \textit{opname},\, \textit{path},\, \textit{tag},\, \textit{type},\, \textit{IP},\, \textit{OP},\, \textit{GC}\rangle$ as described below:

\begin{itemize}[leftmargin=*]
    \item $\textit{opname}$ is a unique name for the operation. This is typically presented in OpenAPI specification as \textit{operationid}. 
    \item $\textit{path}$ is the relative URL for accessing the operation endpoint.
    \item $\textit{tag}$ is an optional set of keywords to tag the operation.
    \item $\textit{type}$ is the HTTP method for RESTful operation that can take values: POST (create), GET (read), PUT (update/replace), PATCH (update/modify), and DELETE (delete).
    \item \textit{IP} is the set of input parameters (top-level or through transitively unrolled schema). Each input parameter $\textit{inp} \in \textit{IP}$ is represented as a 7-tuple $\langle \textit{pname},\, \textit{ptype},\, \textit{is\_required},\, \textit{loc},\, \textit{pc},\, \textit{examples}, \textit{id} \rangle$, where $pname$ is the parameter name, $\textit{ptype}$ is the parameter type which can be either primitive type(such as integer, boolean, string)  or a schema, \textit{is\_required} denotes if the parameter is mandatory, $loc$ can be one of \small{BODY}, \small{PATH}, \small{QUERY}, \small{HEADER}, and  \small{FORMDATA}, and $\textit{pc}$ uniformly captures various constraints at the parameter level, such as range, format (email, ipv4), etc. The following section discusses the supported constraint types. $\textit{examples}$ capture any data values given in the OpenAPI specification (present as `example' or `examples' attribute), to support using domain-specific values. $\textit{id}$ is a unique identifier of the parameter that is a concatenation of the $\textit{opname}$, \textit{loc} and \textit{pname}.
    \item $OP$ is a set of output parameters similar to IP but has $responsecode$ as an additional property for each parameter. $o.responsecode$ denotes the HTTP response code at runtime.
    \item \textit{GC} stands for global constraints such as inter-parameter dependencies~\cite{martin2021specification} and inter-operation producer-consumer dependencies. Such dependencies cannot be formally described in the OpenAPI specification, hence the need for a model to support specifying such constraints. 
    
\end{itemize}
Each schema $s \in S$ is represented as $\langle sname, SF\rangle$ where $sname$ represents the schema name and $SF$ is the set of fields of that schema represented as $\langle \textit{fname}, \textit{ftype}, \textit{is\_required}, \textit{fieldconstraint} \rangle$; these are the same as the components of the input parameter. 

\subsection{Algorithm}
\begin{figure}
    \centering
    \includegraphics[width=5in]{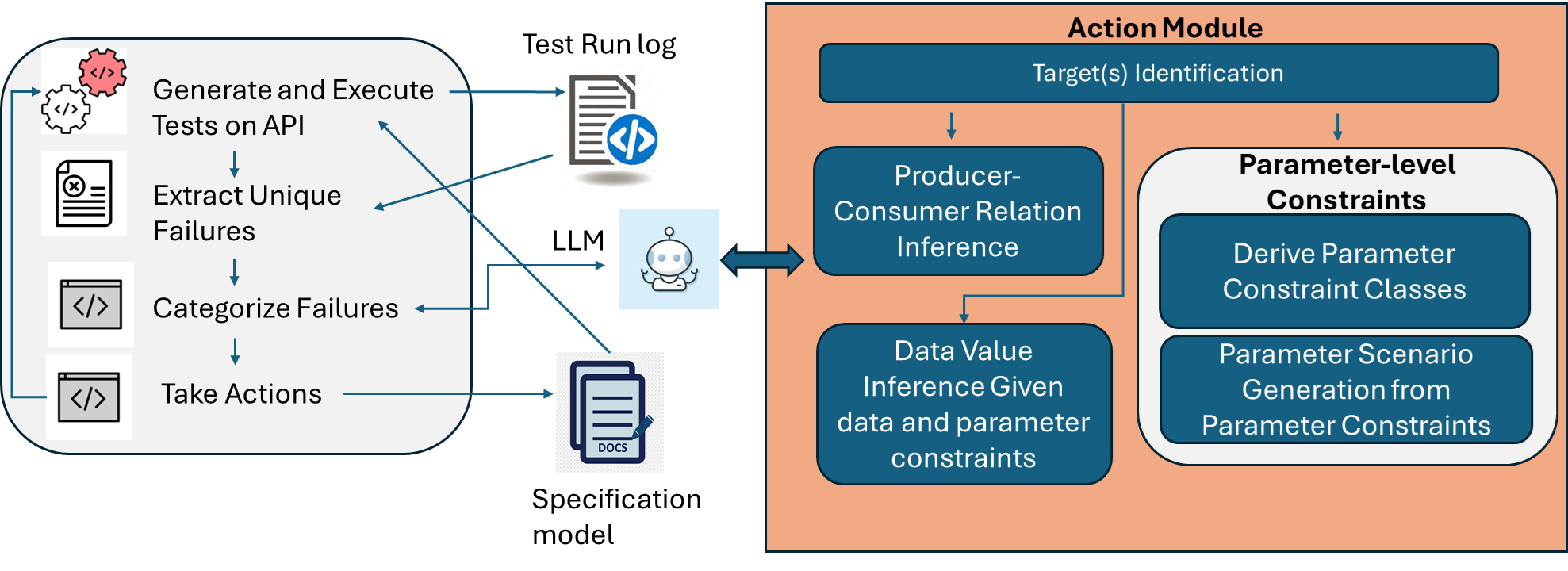}
    \caption{Architecture Design for \system}
    \label{fig:design}
\end{figure}
\eat{
\subsubsection{AI Agent Design}
Agent: API testing agent that invokes appropriate tools to analyze API error messages and take appropriate actions to refine the specification model and run the generated API tests.
}
The functionality of {\system} is driven by the following components, as illustrated in Fig. \ref{fig:design}.
\begin{itemize}
    \item generate and Execute: Use the specification model to generate and run API test cases
    \item Log Failure: Log the list of endpoint operations with unique failure responses.
    \item Categorize Failure: Identify the type of failure based on the response .
    \item Take Action: Depending on the error category, obtain appropriate constraint and update the specification model. This invokes a set of sub-actions (depending on the category). This includes selecting target entity to act on, and if applicable obtain the source entity or data value to finally derive an object of the applicable Constraint class (Table \ref{tbl:classes}).
    
\end{itemize}

Note that the loop-back from \textit{`Take Action'} to \textit{Generate and Execute} represents the iterative-learning nature of the approach.

\begin{algorithm}[t]
\caption{Iterative testing pipeline} 
\label{algo:pipeline}
\SetAlgoNoLine
\footnotesize
\KwData{API\_Spec (JSON file containing the OpenAPI specification), Exec\_params (a dictionary containing the execution parameters such as token to execute the API)}
\KwResult{Refined Specification model, TestRuns dump}
\SetKwProg{Fn}{Function}{}{end}

\Fn{pipeline(API\_Spec, Exec\_params)}{
    // Load and Parse specification into spec\_model \\
    spec\_model := load\_spec(API\_Spec) \\
    learn := True \\
    failures := set() \\
    reports := list() \\
    \While{learn}{
        // Obtain Producer Consumer constraints from the spec\_model \\
        producer\_consumer\_pairs := extract\_dependencies(spec\_model) \tcp*[l]{initially empty list}
        
        // Obtain list of sequences containing operations preceded by pre-requisite operations inferred from producer-consumer dependencies \\
        sequences := generate\_sequences(producer\_consumer\_pairs, spec\_model) \\
        
        // Obtain parameter combinations of each operation - mandatory, all, some optional with mandatory \\
        param\_scenarios := generate\_param\_combinations(spec\_model) \\
        
        // Generate data for each parameter of operations \\
        data := generate\_data(spec\_model) \\
        
        // Generate tests for each sequence, parameter, and data scenario \\
        testcases := generate\_tests(spec\_model, producer\_consumer\_pairs, sequences, param\_scenarios, data) \\
        
        // Run tests to log failures and logs \\
        report, run\_failures := execute\_tests(testcases) \\
        
        // Convergence condition \\
        \If{run\_failures $\subseteq$ failures}{
            learn := False \\
        }
        failures := failures $\cup$ run\_failures \\
        spec\_model := analyze\_failures(spec\_model, run\_failures) \\
        reports.add(report) \\
    }
    \Return{spec\_model, reports}
}

\KwData{spec\_model (Specification model), run\_failures (List of failure responses obtained from test run on API)}
\KwResult{Refined Specification model}

\Fn{analyze\_failures(spec\_model, run\_failures)}{
    \ForAll{failure $\in$ run\_failures}{
        category := classify\_failure(failure) \\
        spec\_model := take\_action(category, spec\_model) \tcp*[l]{returns updated spec\_model with constraints added}
    }
    \Return{spec\_model}
}

\end{algorithm}

Algorithm \ref{algo:pipeline} explains the iterative learning pipeline of {\system}. The algorithm consumes the API specification and additional user configuration to produce a specification model and test cases refined over every iteration of learning from the response feedback. 

Formally, each test case $t$ is represented as a 5-tuple:

$\langle \textit{seq}, \textit{params}, \textit{data}, \textit{checks}, \textit{deps}\rangle$ as described below:

\begin{itemize}[leftmargin=*]
    \item $\textit{seq}$ or sequence scenario is a sequence of operations in the test. 
    \item $\textit{params}$ or the parameter scenario is a mapping between an operation $op$ in the \textit{seq} and a set of input parameters $\subseteq op.IP$.
    \item $\textit{data}$ is the mapping between each parameter in $\textit{param}$ to a value.
    \item $\textit{checks}$ is the set of checks on the response structure of every operation in $\textit{seq}$. A typical check for a valid/positive test case is that the response code should be between 200 and 399.
    \item $\textit{deps}$ is the set of producer-consumer dependencies between the operations in the $\textit{seq}$ where each producer-consumer dependency captures the relationship between input/output parameter of a producer operation with the input parameter of a consumer operation. The test execution engine uses such parameter dependencies to maintain a running state and initialize some input parameters of consumer operations. 
\end{itemize}

The first step of the test case generation process generates a specification model from the OpenAPI specification (Function \textit{load\_spec}). 

\paragraph{Sequence generation} Deriving a valid sequence of operations to execute is important. In every iteration, the latest version of the model is consumed to generate operation sequences to execute later (Function \textit{generate\_sequences}). Each sequence \textit{seq} is essentially a target operation preceded by its pre-requisites.
For example, to successfully execute \texttt{getPetById} operation (target) we need to first execute \texttt{addPet} operation which creates the \texttt{Pet} resource. The algorithm for sequence formation based on produce-consumer dependency is described later. Initially, as the producer-consumer dependency is empty, the initial list of sequences would contain a single operation in each sequence. 

\paragraph{Parameter scenario} To explore diverse combinations of parameters to send in an operation request, we define three types of parameter scenarios. For operations that are pre-requisites, only a single parameter scenario is generated with all mandatory parameters and for target operation multiple scenarios are generated (Function \textit{generate\_param\_combinations}): 1) scenarios with a maximal number of parameters (i.e. all parameters), 2) scenarios with the minimum number of parameters (i.e. mandatory parameters), and 3) minimal length scenarios each covering at least one optional parameter. For instance, if an \textit{All} parameters scenario looks as this: \textit{\{p1, p2, p3\}} and if a constraint exists as \textit{One(p1, p3)}, meaning that only one of p1 and p3 should be present, then instead of All scenario - two scenarios \textit{\{p1, p2\}} and \textit{\{p2, p3\}} are produced. The parameter scenario generation algorithm considering different types of constraints affecting it is described later. 

\paragraph{Data Scenario} For each parameter, various data values are generated, compliant with all the constraints in the specification model (Function \textit{generate\_data}). The data values are randomly generated if no constraint is found on the parameter, otherwise, all the specification constraints related to the selected parameter scenario are considered to generate the data. The algorithm is described later. For each parameter scenario, a predefined number of data scenarios are generated. 

\paragraph{Test Generation and Runs} Test cases are generated as a combination of the operation sequence space, parameter structure space, and data space (Function \textit{generate\_tests}). On execution of the test cases, the run details are logged, along with capturing all \textit{unique} failure responses (Function \textit{execute\_tests}). In this case, {\system} logs all unique responses with 4xx (such as 400, 404) codes to feed to the failure analysis algorithm.

\paragraph{Failure Analysis and Learning} The role of the failure analysis algorithm is to infer the appropriate constraints to add to the specification model, such that the error response could be avoided in the following iterations of the test case generation and execution (Function \textit{analyze\_failures}). This approach aims to obtain valid test cases, especially to minimize the 4xx responses, which are indicative of invalid requests at the API user's end. On the other hand, 5xx responses are unhandled errors at the API's end and usually reveal defects in the API. \eat{However, we observed that often invalid inputs in the request can also lead to 5xx, if not appropriately caught in the API's implementation. As a result, {\system} also analyzes failures giving 5xx response, in addition to logging it as a potential defect.} By learning from the 4xx failures, any subsequent 5xx would be a step closer to revealing defects in the API. The algorithm converges when no \textit{new} failure is logged from the run.

The analysis contains three steps: 1) classify the response to one of the constraint categories, 2) create the constraints by identifying the entities associated with the constraint types, and 3) take action on the identified entities based on the identified constraint category which may involve adding the constraints to the specification model and generate test data with them, ask for user-input typically to obtain authentication/configuration input or ignore. Each of these steps has been described in detail in the following section.

\begin{table}
    \centering
   \scriptsize

    \begin{tabular}{lll}
    \toprule
    \multicolumn{3}{c}{\textbf{API List}}\\
    \midrule
    Car-API & Gestao-Hospitalar & Google-Maps*\footnotesize{\cite{googleGoogleMaps}}\\
    Language-Tool & Market & NCS\\
    Person-Controller & Petstore & Problem-Controller\\ Rest-Countries & RestfulWebService & SCS\\
    User-Management & Xkcd & YoutubeDataApi*\footnotesize{\cite{googleYouTubeData}}\\ YelpBusinessesSearch*\footnotesize{\cite{yelpGettingStarted}} & Stripe*\footnotesize{\cite{stripeErrorCodes}} &\\
    \bottomrule
    \end{tabular}
    
    \caption{\label{tbl:categoryAPIs}APIs used to collect responses to study categories.}
 
    \footnotesize{*Response samples from API documentation were used, limited by the authentication requirements to access the API}
\end{table}

\newcounter{magicrownumbers}
\newcommand\rownumber{\stepcounter{magicrownumbers}\arabic{magicrownumbers}.}
\begin{table*}[]
\footnotesize
    \centering
    \resizebox{\textwidth}{!}{
    
    \begin{tabular}{p{0.5cm} p{3.5cm} p{4.5cm} p{5cm}}
        
    \toprule
    & \textbf{Category} & \textbf{Description} & \textbf{Sample}\\
    \midrule
    \rownumber & Configuration /Authentication & Data is required from user, such as authorization details or credentials & \textit{API key not valid. Please pass a valid API key.}\\
    \midrule
    & \textbf{Operation-level}\\
    \midrule
    \rownumber & ProducerConsumer (producerop, producerparam, consumerop, consumerparam) & A pre-requisite operation exists for the target operation & \textit{playlistNotFound: The playlist with the ID `playlist456' could not be found.}\\
    \midrule
    \rownumber & UnsupportedOperation & Invalid operation present in specification &  \textit{Method Not Allowed Request method `POST' not supported}\\
    \midrule
    & \textbf{Parameter-level}\\
    \midrule

    \rownumber & AdditionalMandatory Parameter (p1.is\_required=true) & Parameter p1 must be present & \textit{"points" is a required parameter.}\\
    \midrule
    \rownumber & Or(p1, p2,.., pn) & If optional parameters exist, at least one should be specified & \textit{You must specify either the `source' or `destination' parameter.}\\
    \midrule
    \rownumber & One(p1, p2,$\ldots$, pn) & Among a subset of parameters only one should be given value & \textit{Either city or zipcode is required, not both.}\\
    \midrule
    \rownumber & AllOrNone(p1, p2,.., pn) & Among a subset of parameters, either all should be given value or none specified. & \textit{Address should be specified with street, city and pincode.}\\
    \midrule
    \rownumber & ConditionalParameter Required 
    (p1, isPresent1, p2, isPresent2) & If param p2 present/absent then param p1 should be present/absent;  & \textit{If longitude specified then latitude should be too}\\
    \midrule
    \rownumber & Parameter-Unknown & When an unknown parameter is present in the request & \textit{Received unknown parameter: url}\\

    \midrule
    & \textbf{Data-level}\\
    \midrule
    \rownumber & Data-Arithmetic (p1, reloperator, p2/const) & p1 is Greater than, less than, equal, unequal to p2 or a constant/list of constants & \textit{afterTimestamp must be greater than beforeTimestamp}\\
    \midrule
        \rownumber & Data-NonArithmetic (p1, property, value) & Invalid data based on type, uniqueness & \textit{`PL' is not a valid gender. Supported values are `Male' , `Female', `Other'.}\\
    \midrule
    & \textbf{Parameter+Data}\\
    \midrule
    
    \rownumber & DataInfluenced-ParamSelection ($DataArithmetic\implies PC$) & Parameter selection constraints based on data values & \textit{If type is 'audio', only one of the other two parameters is required}\\\midrule

        \rownumber & ParameterInfluenced Data Values ($PC\implies DataArithmetic$) & Constraint on Data values if param1 is present/absent  & \textit{If thumbnail is present, type must be `link'.}\\

    \midrule\midrule
    \rownumber & Unhandled & Constraint can't be determined. Should be reported as potential defect & \textit{ Internal Server Error}\\
    \bottomrule
    \end{tabular}
    }
    \caption{\label{tbl:classes}Constraint Categories.}
  
\end{table*}


\subsubsection{Constraint Inference} 

\paragraph{Constraint Categorization}
We conducted a short study to obtain the category of constraints observed in API responses. We collected response messages from over 17 APIs (Table \ref{tbl:categoryAPIs}) when run with API testing tools MOREST \cite{liu2022morest}, EvoMaster \cite{arcuri2021evomaster}. For 4 of these APIs, the tools could not be run due to usage limits. For these particular APIs, we have studied their API documentation to collect response messages to include in the dataset. Additionally, the response data used by Lopez et al. in their analysis work on inter-parameter dependencies \cite{martin2021specification} was also included in the dataset. This resulted in a dataset size of 17,887 response instances. On extracting the unique responses, we obtained a dataset of 1663 responses. These response messages were studied to identify 14 different types of categories. The categories are described in Table \ref{tbl:classes}. 

Except for categories 1, 3, 9, and 14, all other categories are expressed as constraints in the specification model. The OpenAPI does not allow formally specifying categories 5, 6, 7, 8, 12, and 13 and hence, we designed our specification model to accommodate such constraints. Constraints in categories 2-3 impact the sequence scenario formation, categories 4-9 are related to parameter scenario formation and solely dependent on the parameter properties, and categories 10-11 impact the data scenario formation and are solely dependent on data values. Categories 12-13 are the interesting ones as they are conditional to the constraints in other categories. 

\begin{figure}
    \centering
    \includegraphics[width=3in]{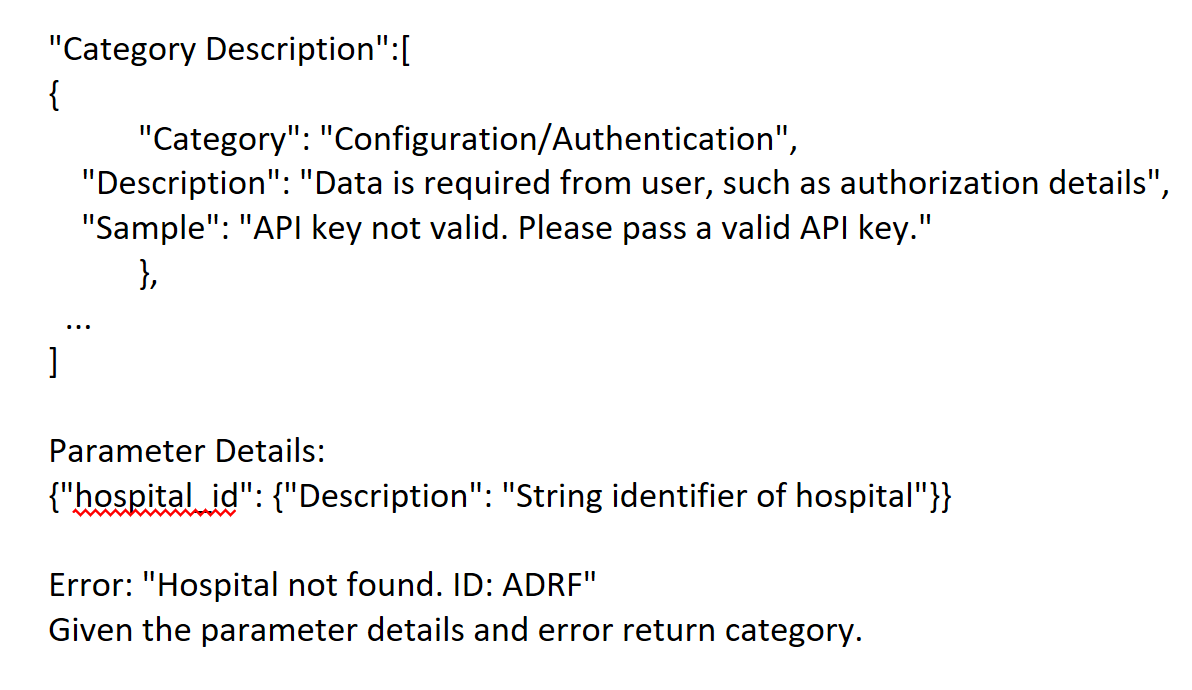}
    \vspace{-4mm}
    \caption{Prompt for categorization}
    \label{fig:prompt}
\end{figure}

Mapping natural language response messages to the above categories is modeled as a classification problem. We have used an LLM to retrieve the constraint types. We use a zero-shot prompt for the Mistral Large-2 model to infer the categories. An example prompt is shown in Figure~\ref{fig:prompt}. When prompting to LLM model cannot identify a category (responds with \textit{`Unhandled'} category), we use an alternate fine-tuned classification model.  We trained a classifier using a BERT-based transformer model \cite{devlin2019bert}, on the response message data to classify a concatenated string of response message and response code to one of the categories described in Table \ref{tbl:classes}. With a train:test split of 80:20, we trained for 5 epochs to obtain the classification accuracy of 95.5\% on the test set. With this, we obtained the classification model to use in {\system}. 

\paragraph{Constraint Identification}

Once the response message is classified into a category, \system{} uses its \texttt{Action Module} to form the specification constraint objects. Its underlying implementation has an intelligent agent based on Mistral-Large model performing tasks such as entity recognition, data generation etc., depending on the constraint category. The end goal of the agent is to derive the constraint object and populate to the specification model at an appropriate point. The format of the specification constraint is shown in Table~\ref{tbl:classes} along with the categories. The $PC$ (parameter constraints) in categories 12 and 13 denotes any constraints from categories 5, 6, 7, and 8. 

Even though we implemented separate functions to form constraints in separate categories, there are two common challenges associated with the constraint formation process: 1) identification of entities associated with the constraints, 2) disambiguation of parameter positions in arithmetic and logical relations, and 3) identification of nested constraints for categories 12 and 13.  

We now discuss the challenges and solutions to infer the constraint entities. These entities could be the API operation, parameter, property of parameters, or constant values. The first problem is to match parameter names with natural language variations. Consider the response message \textit{`Internal Server Error: This email beulalingo@yahoo.com is already in use.'}; from among several candidate parameters \textit{[`gender', `linkedin', `password', `secured', `mobile', `website', `address', `username', `name', `emailId', `country']} for the operation, it should pick \textit{`emailId'} to act on. {\system} infers this entity, using prompt to the LLM by populating it with the dynamically obtained information, such as parameter names of the operation and error message for the given example.

Consider the expected constraint ProducerConsumer(producerop=\textit{addPet}, producerparam=\textit{id}, consumerop=\textit{createOrder}, consumerparam=\textit{petid}) and a relevant message `Pet not found' from PetStore benchmark while testing \textit{createOrder} operation (POST operation) for creating an order related to a pet. The AI agent consumes information about the 1) identified target (consumer parameter) from the response message and 2) list of potential producer operations (POST type operations) and their parameters to infer the ProducerConsumer pair. With the suitable intelligence, the agent identifies the producer as the post-operation which is closely associated with the noun entity (Pet) in the message. Producer's id-output param  becomes the \textit{producerparam}. The consumer's input param \textit{petId}, which is closely associated with the combination of the producer's resource name (Pet) and producer parameter name (id) is considered as \textit{consumerparam}.  

Data-level constraints are similarly handled by the agent by consuming the target and the response error to producer set of realistic values to use as test data.

The second challenge pertains to identifying and encoding logical relations between parameters from natural language. For instance, an error message: \textit{`gain should surpass expenditure'} can be expressed logically as \textit{gain $>$ expenditure}. The latter is easier to analyze to generate data values when encoded as constraint ConditionalParameterRequired(p1= \textit{gain}, reloperator= \textit{$>$}, p2= \textit{expenditure}). To obtain such an expression, the underlying agent consumes the response and the list of identified target parameters.

To identify the nested constraints in categories 12 and 13, we break the sentence into two parts corresponding to the antecedent and consequence of the implication. To identify the $PC$ part of the constraints, we use the relevant part and follow the classification procedure to classify the type of parameter constraint and then use the relevant constraint formation procedure for specific parameter constraint categories.

\paragraph{Actions and Scenario Formation}

Once all the categories are identified with relevant constraints, there are different types of actions that can be taken based on the categories.  For the Configuration/Authentication (1) category user is asked to provide the configuration details.  The unsupported operations (3) and unknown parameters (9) are deleted from the specification model. Unhandled cases (14) are reported as errors and no further processing is done. The rest of the cases influence the sequence, parameter, and data scenario formation, and hence we describe how such formation handles the multitude of inferred constraints.\\

\noindent{\bf Sequence Scenario.} As stated earlier sequence scenario generation uses producer-consumer dependencies to form pre-requisite for the target operation. For each operation (target), we identify its prerequisites by transitively traversing the producer-consumer dependency with consumers as known. If for an operation multiple producers are found then only a single producer is used which generates a single resource. Note that, we have seen multiple producers are possible for a resource/schema when one producer generates one resource and another producer generates multiple resources of the same type. For testing a consumer operation, not all its producers are required. All the operations are topologically ordered.      \\

\noindent{\bf Parameter Scenario.} 
The parameter scenario generation algorithm expresses all the parameter selection-related constraints into Z3-constraints. Each parameter gets either 1 or 0 value correlating whether it will be selected or not. All mandatory parameters get 1 value. If any optional parameter of a schema is 1 then the mandatory parameter of the schema is also 1. If all optional parameter list is $OL$, then $Or(p_1,\ldots\,p_n)$ constraint is expressed as of $sum(p \in OL)>0 \implies sum(p_1,\ldots,p_n)>0$. $One(p_1,\ldots,p_n)$ is expressed as: $sum(p_1,\ldots,p_n)>0 \implies sum(p_1,\ldots,p_n)=1$. $AllOrNone(p_1,\ldots,p_n)$ constraint is expressed as $sum(p_1,\ldots,p_n)=0 \vee sum(p_1,\ldots,p_n)=n$. $ConditionalParameterRequired(p1,b1,p2,b2)$ is expressed as $p2=int(b2) \implies p1=int(b1)$. The challenging case is DataInfluenced-ParamSelection. Since parameter scenarios are generated before data scenario generation, we simply divide this constraint to create two constraints: one where all parameters related to DataArithmetic parameter are present (i.e. adds up to its size) along with the existing PC constraint, and another case where this constraint is omitted. 

The algorithm repeatedly calls constraint solver to obtain different parameter scenarios and then filters to get parameter scenarios with maximal length, and minimal length, and iteratively finds a minimal length scenario covering each optional parameter. \\

\noindent{\bf Data Scenario.} Given a parameter scenario, we first gather all the relevant data constraints containing all parameters in the scenario. If $ParameterInfluencedDataValues(\textit{PC}\implies \textit{DataArithmetic})$ exists, then $PC$ is evaluated against the parameter scenario, and if it holds then \textit{DataArithmetic} constraint is also gathered.  If $DataInfluenced-ParamSelection(DataArithmetic \implies PC)$ is present, \textit{DataArithmetic} constraint is also gathered if $PC$ evaluates to true. To generate realistic values satisfying all the constraints, we create a prompt expressing all the constraints in natural language and query llama-2-7b model \cite{touvron2023llama} to generate the required number of data scenarios. The resultant scenarios are checked against the gathered data constraints (using a Z3 constraint solver) and if not satisfied then Z3 solver is used to generate values satisfying the data constraints. Otherwise, realistic values satisfying the data constraints are used for test case generation.

\eat{
\paragraph{Rule to Constraint Mapping}
Once the category of constraint and the entity to be acted upon is derived, the next task is to create the constraint object containing the related details. Further, action classes are designed on top of these constraint classes to reflect in various components in the {\system}. The action classes are listed below for every constraint, describing the action taken.
\begin{enumerate}
     \item \textit{UserData}- The user is expected to pass the data values either in the execution configuration or the specification.
    \item \textit{DataConstraint}- {\system} handles this constraint using LLM to generate a sample value and constraints for a given response error. For a given error message, {\system} generates prompt to llama-2-7b model \cite{touvron2023llama} seeking a JSON object containing the target parameter, sample value and data value constraint in the form of regular expression (if string type) or minimum, maximum range (if numeric type). The response is parsed to create Z3 compatible constraint object to the specification model and example value.
    \item \textit{ProducerConsumer}- The consumer resource candidate, which is the target parameter in the response error, is identified to fetch the producer operation falling in the same grouping as the consumer operation, based on the resource schema on which it operates. For instance, \texttt{getUser} and \texttt{addUser} are operation on the User schema and will be grouped together. If no candidate exists in the same group, then other groups are explored. For every identified producer candidate operation, their response parameter names are matched with that of the target parameter to get the closest match for the produced resource. Hence, the derived tuple (Prod\_op,Prod\_rsrc,Cons\_op,Cons\_rsrc) is added in the specification model to the list of constraints on the consumer operation that gave the response error. These constraints are consumed when inferring the producer-consumer dependency in the API testing cycle.
    \item \textit{UnsupportedOperation}- Such operations are deleted from the model and, hence not included in the generated test sequences.
    \item \textit{ParameterRequired}- The required tag of the inferred target parameter is set to True. Hence, when generating the parameter scenarios, the parameter is treated as mandatory.
    \item \textit{ConditionalParameterRequired}- We prepare a fine-tuned entity recognition model on top of flan-t5-xl(3B) model \cite{flan} for a given response error and list of parameters under consideration such that the entities recognised are the tuple (independent\_parameter, independent\_parameter\_datavalue, dependent\_parameter, dependent\_parameter\_datavalue). For instance, if the error says "if p1 is set to `auto', p2 should be `english'", the tuple predicted would look as (p1, `auto', p2, 'english'). The learnt constraint is added to the list of constraints for the concerned operation and the test scenarios are generated based on this.
    \item \textit{Or(p1, p2,.., pn)}- The parameters in this `Or' set are added one at a time, to the existing parameter scenarios, if not present, hence, generating updated parameter scenarios where at least one of these parameters are present in each scenario. Note that for scalability sake we do not generate all possible scenarios for this constraint, such as taking pair-wise(or triplet etc.) combination of parameters in the `Or' set.
    \item \textit{One(p1, p2,.., pn)}- The parameters in this `One' set are added one at a time, to the existing parameter scenarios, if not present, hence, generating updated parameter scenarios where exactly one of these parameters are present in each scenario.
    \item \textit{AllOrNone(p1, p2,.., pn)}- Parameter scenarios are updated such that either all mentioned parameters are present in the parameter scenario or none, to the existing scenarios.
    \item \textit{Parameter-Arithmetic}-  We prepare a fine-tuned entity recognition model on top of flan-t5-xl(3B) model \cite{flan} for a given response error and list of parameters under consideration such that the entities recognized are the tuple (par1, par2, operator). par1 and par 2 are LHS and RHS operands with supported operands as $<, <=, >=,$ and $=$.
    \item \textit{Parameter-Complex}- The empirical instances observed for such cases have been supported under the action taken for \textit{ConditionalParameterRequired} category. We shall support a generic approach in future for this category.
    \item \textit{Parameter-Unknown}- Such parameters are removed from the specification, and hence from test scenarios.
    \item \textit{Unhandled}- Logged as a failure and no action is taken.
\end{enumerate}
}

\paragraph{Handling Blank Response}
{To handle scenarios of blank response messages, which were found in ~22\% of the requests over the benchmark APIs, {\system} additionally leverages the response code to infer possibility of a 4xx. It derives heuristics to handle such responses. For instance, 404 response is  a \textit{`Not found'} scenario and if the operation under test is found to be requiring and identifier in its input, then the case is treated to be a ProducerConsumer Constraint. The LLM is responsible to identify if any of the input parameters represent an identifier. Other 4xx blank responses are assumed to be a data issue and the subsequent data values for such operations are generated by the agent. Applying these heuristics, {\system} was able to resolve several 4xx responses even in the absence of a response message. Nevertheless, the content of the response is certainly more useful in making the correct refinement to the test case.}
\section{Experimental Evaluation}
\label{sec:expt}

This section presents the evaluation of {\system} to investigate the following research questions.
\subsection{Research Questions}
\begin{itemize}
    \item \textit{RQ1}: How effective is {\system} with respect to coverage metrics?
    \item \textit{RQ2}: On effectiveness of learning in generating functional test cases
        \begin{itemize}
            \item \textit{2a} How effective is {\system} in reducing invalid test cases with every iterative learning?
            \item \textit{2b} How effective is {\system} in generating valid test cases? 
        \end{itemize}
    \item \textit{RQ3}: How does {\system} perform with respect to operation hits made, as compared with the state of the art tools? 
    \item \textit{RQ4}: How effective is {\system} in revealing defects in the API? 
\end{itemize}
\begin{table}[t]
\scriptsize

    \caption{Benchmark services used in the evaluation.}
    \label{tab:bench}
    \centering
    \begin{tabular}{ccc}\toprule
    \textbf{API} &     \textbf{Operations} &  \textbf{GET, POST, PUT, DEL}\\
    \midrule
    petstore~\cite{petstore} &  20 &  8, 7, 2, 3 \\    
    person~\cite{restgo} & 12 & 5, 2, 2, 3 \\
    genome~\cite{restgo} & 23 & 13, 10, 1, 0\\
    gestao~\cite{gestao} & 20 &  10, 5, 3, 2\\
    user mgmt.~\cite{restgo} & 23 &  6, 7, 2, 5\\
    market~\cite{restgo} & 13 &  7, 2, 3, 1 \\
    problem cont.~\cite{restgo} & 8 &  2, 3, 1, 2 \\
     langtool~\cite{languagetool} & 2 &  1, 1, 0 , 0 \\
    scs~\cite{restgo} & 11 &  11, 0, 0, 0 \\
    ncs~\cite{restgo} & 6 &  6, 0, 0, 0\\
    rest countries~\cite{restgo} & 22 &  22, 0, 0, 0\\
\bottomrule
    \end{tabular}
\end{table}

\subsection{Experimental Setup}
All experiments were run on a machine with Microsoft Windows 11 OS, with 32 GB RAM and 11th Gen Intel Core i7-1165G7 2.80GHz processor. {\system} is implemented in Python.
\subsubsection{Benchmarks}
We selected 11 public APIs, that have been used in the past research (\cite{kim2023adaptive, restgo, arcuri2018evomaster}). These 11 APIs were shortlisted applying a similar elimination strategy as was applied for ARAT-RL\cite{kim2023adaptive}. We eliminated APIs that required authentication, had external dependencies on services or had usage rate limits. We used these as benchmarks to evaluate {\system}. Table~\ref{tab:bench} lists the benchmarks, along with details of number of operations and distribution of operations over various REST calls. 

\subsubsection{Tools}
To investigate the first three RQs, we compare {\system}  with three state-of-the-art API testing tools, ARAT-RL~\cite{kim2023adaptive}, EvoMaster (v1.6.0) and MOREST~\cite{liu2022morest}\cite{arcuri2021evomaster}. These three tools have been picked as they have been shown to be the best performing API testing tools in the recent paper by Kim et al.~\cite{kim2023adaptive}. Considering, {\system} is a blackbox tool, we use the blackbox mode of EvoMaster for a fair comparison. These tools have been executed with their default configurations, setting a time budget to 30 minutes, wherever applicable. {\system} does not have a time budget as its termination condition is when no error or previously seen errors occur in the next iteration. Note that the time budget set for the other tools is more than the maximum of 16 minutes (average 5 minutes) run time seen over one of the benchmarks by {\system}.  Furthermore, for a fair evaluation of {\system}, that focuses on the effectiveness of  API testing with minimal requests, we also executed ARAT-RL, EvoMaster and MOREST with an operation hit budget of the number of requests made by {\system} for each benchmark API. This budget has been obtained from averaging over 3 runs of {\system} on each benchmark API. To summarise, each of the three benchmarks tools has been executed in two modes: 1) time budget of 30 minutes, and 2) hit limit \textit{(hl)} of requests made by {\system} for the corresponding benchmark API. The results presented have been averaged on 3 runs of each tool.

The code of MOREST and ARAT-RL was altered by adding hit limit constraint to the loops. For EvoMaster, the \texttt{maxActionEvaluations} and \texttt{stoppingCriterion} parameters were passed to impose this limit, as suggested by one of the authors of the tool.

\subsubsection{Result Collection}
Data collection for code coverage was done using JaCoCo\cite{eclemmaEclEmmaJaCoCo}. Other metrics of operation coverage, 2xx, 4xx, 5xx response were extracted from the output reports produced by the tools, using a python script we wrote for each of the tools. The results are available as supplementary material.

\subsection{Results}

MOREST could not be executed on the \textit{user} API as it terminated with a 'keyError'. ARAT-RL could not be executed on \textit{problem} API as the specification parsing failed due to recursive schema references. EvoMaster and {\system} were successfully executed on all APIs in the benchmark. Table \ref{tbl:rq3} indicates the total number of requests made by each tool, without a hit limit, over benchmarks.

\begin{table}[ht]
\scriptsize
\centering
\caption{\small Operation hits across tools.}
\begin{tabular}{l|r|r|r|r}\toprule
\footnotesize
\textbf{API} & {\textbf{{\system}}} & {\textbf{MOREST}} & {\textbf{EvoMaster}} & {\textbf{ARAT-RL}} \\
\midrule 
petstore & 115 & 7880 & 50  & 41,600\\
\midrule
person & 42 & 46,199 & 23 & 53,043\\
\midrule
genome & 157 & 1947 &  50 & 17,631\\
\midrule
gestao & 186 & 5903 & 60 & 16,800\\
\midrule
user & 134 & - & 60 & 44,577\\
\midrule
market & 132 & 19,676 & 26 & 19,536\\
\midrule
problem & 68 & 7156 & 19 & - \\
\midrule
langtool & 37 & 20 & 2 & 247\\
\midrule
scs & 11 & 64,216 & 12 & 72,148\\
\midrule
ncs & 6 & 31,059 & 16 & 89,888\\
\midrule
rest countries & 37 & 4620 & 25 & 73,734\\\bottomrule
& 925 & 188,676 & 343 & 429,204\\
\bottomrule
\end{tabular}
\label{tbl:rq3}
\vspace{-2mm}
\end{table}

\begin{figure*}[th]
\small
\centering
\includegraphics*[scale=0.4,trim={0 0 0 0},clip=true]
{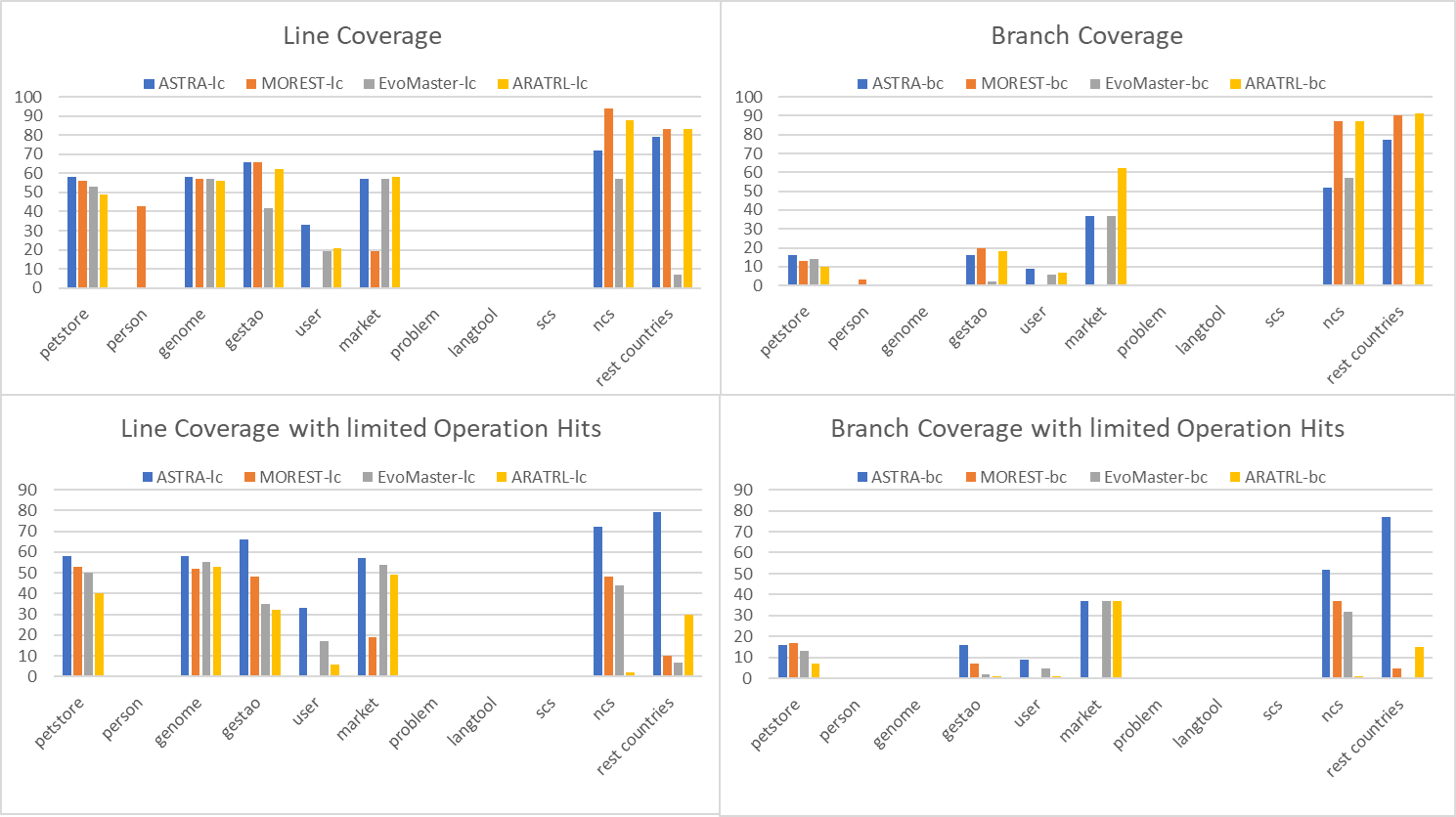}
\caption{Line and Branch Coverage of 4 tools across 11 APIs, with and without OpHit Limit
}
\label{fig:cov}
\end{figure*}
\begin{table*}[t]
\scriptsize
\centering
\caption{\small Operation Coverage Metrics with and without hit limit budget; values in percentage.}
\resizebox{\textwidth}{!}{
\begin{tabular}
{l|rr|rrrr|rrrr|rrrr}\toprule
\textbf{API} & \multicolumn{2
}{c|}{\textbf{{\system}}} & \multicolumn{4}{c|}{\textbf{MOREST}} & \multicolumn{4}{c|}{\textbf{EvoMaster}} & \multicolumn{4}{c}{\textbf{ARAT-RL}} \\
\cmidrule{2-15}
    & oc & $oc_{2xx}$ & oc & $oc_{2xx}$ & $oc_{hl}$ & $oc_{hl2xx}$ & oc & $oc_{2xx}$ & $oc_{hl}$ & $oc_{hl2xx}$ & oc & $oc_{2xx}$ & $oc_{hl}$ & $oc_{hl2xx}$ \\\midrule 
    petstore & \textbf{100} & 90 & 75 & 75 & 20 & 15 & 75 & 55 & 60 & 55 & 50 & 50 & 20 & 15\\
\midrule
person & \textbf{100} & 25 & 92 & 58 & 33.3 & 0 & 83.3 & 25 & 83.3 & 25 & 83.3 & 58.3 & 17 & 8.3 \\
\midrule
genome & \textbf{100} & 100 & 96 & 91.3 & 35 & 35 & 65 & 65 & 61 & 61 & 96 & 96 & 56.5 & 56.5\\
\midrule
gestao & \textbf{90} & 85 & 90 & 90 & 40 & 35 & 25 & 25 & 0 & 0 & 80 & 80 & 20 & 15 \\
\midrule
user & \textbf{87} & 61 & - & - & - & - & 52 & 30 & 39 & 17.4 & 74 & 52.2 & 30.4 & 13\\
\midrule
market & 54 & 15.4 & 100 & 0 & 85 & 0 & 38 & 15.4 & 31 & 8 & 62 & 15.4 & 23.1 & 0 \\
\midrule
problem & 50 & 38 & 63 & 38 & 50 & 25 & 63 & 38 & 63 & 38 & - & - & - & -  \\
\midrule
langtool & \textbf{100} &  100 & 50 & 50 & 50 & 50 & 50 & 50 & 50 & 50 & 100 & 100 & 50 & 50 \\
\midrule
scs & \textbf{100} & 100 & 100 & 100 & 9.1 & 9.1 & 100 & 100 & 91 & 91 & 100 & 100 & 27.3 & 27.3 \\
\midrule
ncs & \textbf{100} & 100 & 100 & 100 & 100 & 100 & 100 & 100 & 83.3 & 83.3 & 100 & 100 & 0 & 0\\
\midrule
rest countries & \textbf{100} & 100 & 100 & 100 & 0 & 0 & 0 & 0 & 0 & 0 & 100 & 100 & 9.1 & 9.1  \\\bottomrule

\end{tabular}
}
\label{tbl:rq1oclimit}

\end{table*}
\subsubsection{\textbf{RQ1: Performance on Coverage Metrics}}
We computed 4 coverage metrics for the 4 tools across the benchmark APIs- line coverage, branch coverage, operation coverage and successful operation coverage. The line coverage (\textit{lc}) and branch coverage (\textit{bc}) have the standard meaning. Operation coverage (\textit{oc}) measures the percentage of operations successfully executed, with 2xx or 5xx response, with respect to the total operations in the API specification. We further computed a successful operation coverage (\textit{$oc_{2xx}$}) that measures the percentage of operations successfully executed, with 2xx response, with respect to the total operations in the API specification. Except for {\system}, additional metrics, \textit{$oc_{hl}$} and \textit{$oc_{hl2xx}$}, were computed for the remaining three tools, representing the two types of operation coverage observed when the hit limit budget was imposed.
For 4 APIs, Jacoco results showed a 0\% line and branch coverage. This seemed incorrect as responses from the API were observed. Despite debugging, we were unable to resolve this and hence, indicating these metrics as unknown for \textit{person, problem, langtool} and \textit{scs}.

Fig. \ref{fig:cov} shows the line and branch coverage trends for all the 4 tools. Over the 7 APIs for which the line coverage could be obtained for all 4 tools, while {\system} outperforms for 3 APIs- petstore, genome and user, it was observed at par with the best performing tools for 2 APIs- gestao and market. It did not perform the best on the branch coverage metric; this could be attributed to the reason that {\system} performs a focused test generation, driven by response-feedback, to reduce 4xx responses. As a result, it may not generate diverse data for operations covered successfully, as it is better suited for directing towards functional tests. However, with the hit limit, {\system} emerges as the strongest, indicating its efficacy in learning with minimal data seen from the responses. 

\eat{{\system} performs a response-feedback driven focused test generation approach, aimed at reducing 4xx responses. Due to this design, it may not generate diverse data for operations covered successfully, as it is better suited for directing towards functional tests. As a consequence, operations that have a pre-requisite operation, may get covered. As result, the line and branch coverage may not increase significantly corresponding to a particular operation, while due to operations getting reached by resolving the pre-requisites, more operations are likely to get covered.}

Table \ref{tbl:rq1oclimit} presents the operation coverage metrics in percentage, as computed over 11 benchmarks for the 4 tools.
It indicates that without imposing a hit limit on the benchmark tools, {\system} outperforms or is at par with them on \textit{oc} for 9 APIs. The reason why it missed certain operations on the remaining 2 APIs is that other tools were able to obtain a 5xx response on the operations missed by {\system} through random data and invalid parameter scenarios. The focus of {\system} is to drive tests towards valid input in order to highlight more meaningful potential defects, it minimizes invalid tests and hence, misses on these operations resulting in a 5xx. To fairly evaluate on how many of the operations were covered through valid tests, we also captured coverage of operations resulting in a 2xx response. {\system} also performs the best over \textit{$oc_{2xx}$} for 9 APIs.  An interesting case of market API indicates that while \textit{oc} is highest for MOREST, it has the lowest \textit{$oc_{2xx}$}, indicating all 5xx responses obtained by it. On investigating, all the error messages indicated ``internal server error", which could be attributed to the random nature of data values. For market API, \textit{$oc_{2xx}$} for {\system} was seen to be at par with the best performing tool.

As evident from Table \ref{tbl:rq3}, MOREST and ARAT-RL make significantly more requests than {\system}. With thousands of hits being made by them, this can be a practical obstacle for industry APIs owing to quota limits and computational expense, which incurs a cost per API call to the user \cite{gamez2019role}. Hence, we evaluated how quick these tools are in exploring the search space with a hit limit. As indicated by the \textit{$oc_{hl}$} and \textit{$oc_{hl2xx}$} in Table \ref{tbl:rq1oclimit}, {\system} emerged as a clear winner, with its $oc$ being the highest for 10 of the 11 APIs and $oc_{2xx}$ being the highest for all APIs. The operation coverage metrics fell sharply, especially for MOREST and ARAT-RL, on applying this budget. This is indicative of reinforcement learning-based approach being impractical for industry API testing.

To summarize, {\system} achieves the best coverage with practically feasible number of requests. This is attributed to its feedback-driven greedy approach that learns faster from limited responses.

\subsubsection{\textbf{RQ2: Effectiveness of Learning}}
\begin{figure*}[th]
\small
\centering
\includegraphics*[scale=0.55,trim={0 0 0 0},clip=true]
{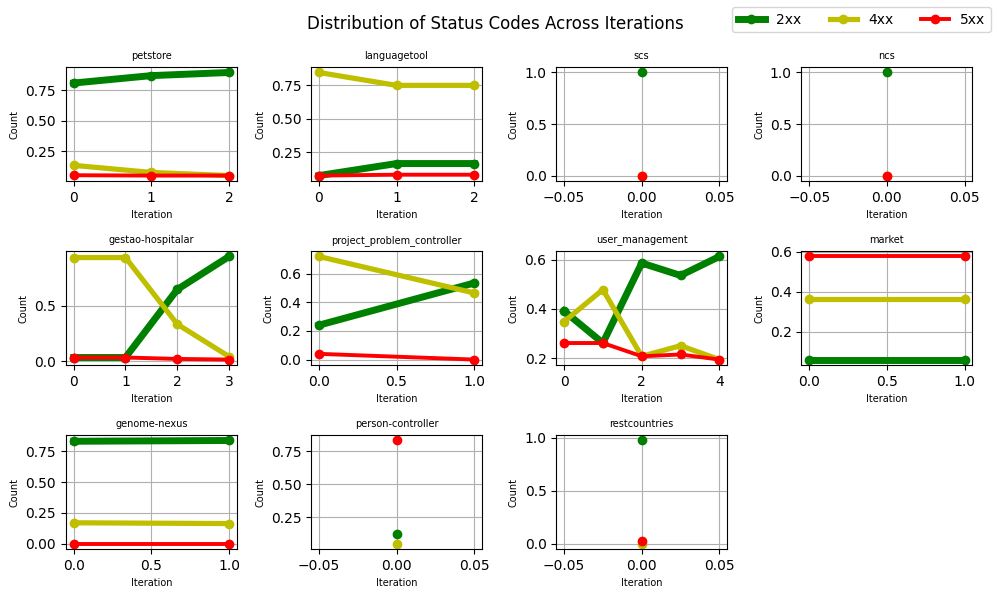}
\caption{Trend of Response Code with each Iteration of Refinement with \system}
\label{fig:iter}
\vspace{-4mm}
\end{figure*}

\begin{figure}[th]
\small
\centering
\includegraphics*[scale=0.4,trim={0 0 0 0},clip=true]
{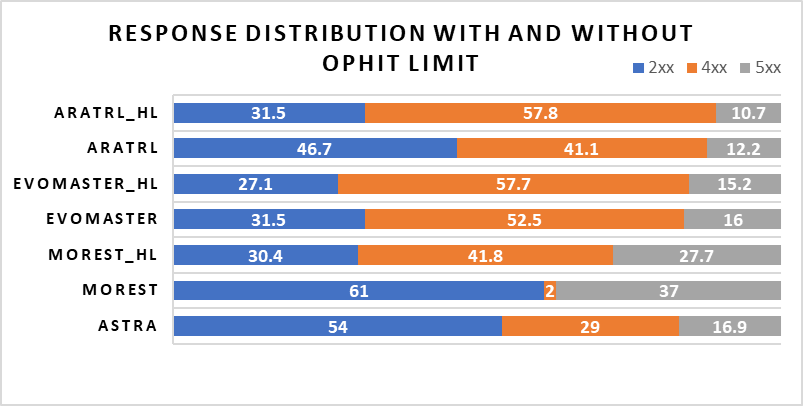}
\caption{Response Code Distribution
}
\label{fig:resp_dis}
\end{figure}

Fig. \ref{fig:iter} indicates how the proportion of 2xx, 4xx, 5xx responses over total responses change with every iteration of {\system}. The overall trend shows that the 2xx, 5xx increase, while, 4xx responses decrease. The change seems significant for 5 APIs (petstore, language, gestao, problem and user) and no change for 6 (scs, ncs, market, genome, person and restcountries). Gestao and petstore contain several producer-consumer dependencies which are learnt from the response with each iteration. \textit{langtool} contained various inter-parameter dependencies and data-value constraints that are learnt from response messages to refine the parameter scenarios with each iteration. 

Two APIs, \textit{market} and \textit{person}, resulted in 5xx response covering all operation in the first run. \textit{scs}, \textit{ncs}, \textit{genome} and \textit{restcountries} covered all operations with 2xx responses in the first run, hence, deriving no learning.

To summarize, the overall observations show 1) a reduction in invalid test cases and 2) an increase in valid test cases, indicating the effectiveness of learning from response messages in refining the test cases for functional testing.

\subsubsection{\textbf{RQ3: Operation Hits}} The ideal case would be to make minimal operation hits, while maximizing the coverage and defect revelations, especially in an industry API that may be performing computationally intensive services.
Fig. \ref{fig:resp_dis} shows the proportion of various response status received across the benchmarks by all 4 tools, without and with hit limit(suffixed as HL).

Even with a much lesser operation hits as compared with MOREST and ARAT-RL, {\system} was able to produce better proportion of 2xx responses than ARAT-RL and EvoMaster, while not very far behind MOREST. With \textit{HL} mode, {\system} defeats all 3 tools. With respect to the minimal proportion of 4xx responses, a similar performance rank was observed. Accounting for 2 APIs \textit{(user, problem)} that could not be executed by some benchmark tools, we omitted these 3 APIs and recomputed the numbers that can be found in the supplementary material. The same trend was observed.

\subsubsection{\textbf{RQ4: Defect Detected}}
\begin{table}[t]
\scriptsize
    \caption{Unique Defects detected by {\system} and MOREST.}
    \label{tbl:rq4}
    \centering
    \begin{tabular}{cccc}\toprule
    \footnotesize
    \textbf{API} &    \textbf{\# Defects by \system} & \multicolumn{2
}{c}{\textbf{{\# Defects by MOREST}}} \\
    & & w/o Hit budget & With Hit budget\\
    \midrule
    petstore & 3 & 0 & 1\\    
    person &  11 & 10 & 4\\
    genome & 0 & 1 & 0 \\
    gestao & 1 & 2 & 2 \\
    user & 6 & 0 & 0\\
    market & 8 & 13 & 11  \\
    problem &  1 & 2 & 2\\
    langtool & 2  & 0 & 0\\
    scs & 0 & 0  & 0\\
    ncs &  0 & 0 & 0\\
    rest countries & 1 & 1 & 0\\
    \bottomrule
    & 33 & 29 & 20\\
\bottomrule
    \end{tabular}
    
\end{table}

To evaluate the potential of {\system} in revealing defects, we gathered the count of unique 5xx responses obtained by it across APIs. As MOREST received the maximum proportion of 5xx responses, we select MOREST to compare our defect results against. Table \ref{tbl:rq4} indicates the unique defects obtained across the benchmark operations by both tools, with and without the hit limit applied. {\system} revealed more number of defects than MOREST, in both the settings. Of the 33 defects revealed by  {\system}, 13 contained the stack-trace, which can be deemed to be serious in nature. MOREST revealed 10 such defects containing stack-trace. In our future experiments, we would be interested in removing the termination condition on {\system} to continue running for a time budget to explore if the defect revealing capability improves. The current numbers are, nonetheless, indicative of the effectiveness, as the defect revealing 5xx responses were observed to be better.

\textit{To conclude the findings from the evaluation, {\system} was observed to perform the best on coverage metrics and in generating valid test cases, even with limited API requests. The iterative approach seems to be effective in reducing the invalid test cases, taking the testing technique closer to exploring defects in the APIs. These results make it suitable for industry usage.}

\subsection{Threats to Validity}

While the results on {\system} have been computed with the termination criteria enabled, a larger time budget was set for other tools. This poses an internal threat to validity. The presented approach aims to perform test refinement with optimizing the operation hits; however, taking liberty on making a higher number of operation hits may also end up revealing more defects in APIs and a better line and branch coverage. Nevertheless, even with the request-conservative, approach we see promising results that are better or at par with benchmark tools.

External threats may come from the choice of LLM model for certain components of {\system}, such as classification and data constraint learning. For every respective component requiring LLM, the choice of model was made from among SOTA openly available models giving stable and best results, as assessed on 3 APIs randomly picked from the benchmark with over 3 runs of {\system} on each. The metrics along each run remained in close proximity. As in the future, improved models may be released, this would only complement the performance of {\system}.

The results may not generalize to other APIs, posing an external threat to validity. While a diverse set of APIs was picked from the past literature on API Testing, it would be of our interest to evaluate {\system} over industry APIs. The list of classification categories may not be exhaustive, however, we could map each failure in the benchmark to one of the defined categories.

\section{Related Work}
\label{sec:related}


Rest API Testing techniques can be broadly classified into black-box and white-box. As our technique is black-box, below we focus on black-box API testing papers.

Atlidakis et al.~\cite{atlidakis2019restler, atlidakistesting} created RESTler, which creates operation sequences from the producer-consumer relationship and explores the operations in a breadth-first fashion. Their sequences do not maintain any resource life-cycle relationship. As identified in ~\cite{liu2022morest}, their producer-relationship contains many false positives. The authors extended RESTler with intelligent data fuzzing \cite{godefroid2020intelligent}. \eat{They employ structural schema fuzzing rules, perform various rule combinations, search heuristics, extract data values from examples including API specifications, and learn data values on-the-fly from previous service responses. Their aim to find 5xx errors, which they extended in \cite{atlidakis2020checking} for checking security-related properties.}

Arcuri et al. \cite{mcminn2004search, arcuri2017many,arcuri2017restful, arcuri2018evomaster,  arcuri2018test, arcuri2019restful,  arcuri2020automated, arcuri2021enhancing, arcuri2021evomaster} 
 developed EvoMaster which includes novel techniques for test case generation for API with white-box access using a genetic algorithm. They also have a black-box version of the algorithm~\cite{arcuri2020automated} which essentially performs random testing adding heuristics to maximize HTTP response code coverage. 

RestTestGen~\cite{viglianisi2020resttestgen,corradini2022automated} infers parameter dependencies with better field name matching than just plain field-name equality which uses field names and object names inferred from operations and schema names. They maintain the CRUD order and create a single sequence to test all the operations for \emph{all} resources in the API. \eat{It does not explicitly mention its strategy for exploring parameter space.} It uses a response dictionary and uses examples and constraints in API spec and random values if such information is not available.

bBOXRT~\cite{laranjeiro2021black} performs robustness testing of APIs. 
Even though this is a black-box tester, it performs negative testing and therefore is not comparable to our approach.  

Giamattei et al.~\cite{russo2022assessing} proposed a black-box testing approach called uTest which uses a combinatorial testing strategy~\cite{nie2011survey}. The input partition for all parameters is formed based on \cite{bertolino2020devopret} and divided into valid and invalid classes.  It uses examples and default values from specifications and bounds to generate random values. They perform test case formation for each operation. No details about the producer-consumer relation or sequence formation are mentioned.

RESTest~\cite{martin2021restest} uses fuzzing, adaptive random testing, and constraint-based testing where constraints are specified manually in an OpenApi extension language, especially containing the inter-parameter dependency. Their sequence generation is confined to each operation with the required pre-requisite. 

Corradini et al.~\cite{corradini2021restats} implemented the black-box test coverage metrics defined in~\cite{martin2019test} in their tool called Restats and performs operation-wise test case generation. 

Tsai et al.~\cite{tsai2021rest} use test coverage ~\cite{martin2019test} as feedback and guide the black-box fuzzers in their tool called HsuanFuzz. They do not produce resource-based functional test cases but follow the order of test operations based on paths and CRUD order. They also use the concept of ID parameters. 

MOREST~\cite{liu2022morest} performs black-box API testing where it first captures the following dependencies - 1) operation $o$ returns an object of schema $s$,  2) operation $o$ uses a field or whole of the schema $s$, 3) operation $o_1$ and $o_2$ are in the same API 4) two schema $s_1$, $s_2$ which has a field in common in the form of resource-dependency property graph. It uses the graph to generate the test cases and rectify the property graph online. 

ARAT-RL~\cite{kim2023adaptive} is another black-box API testing that uses a reinforcement learning-based approach to incorporate response feedback to prioritize the operations and parameters in its exploration strategy.
The approach aims to maximize coverage and fault revelation. While the approach uses the response status code to update the reward and response data to derive sample data values, it misses processing the information-rich error messages to incorporate into its search approach.
\eat{
A recent approach, called RestCT~\cite{wu2022combinatorial},  uses a sequence generation algorithm that considers the CRUD partial ordering constraints and producer-consumer relationship (same as ours) but differs in the notion of resource-boundary-based generation or language transformer-based total ordering, instead generates all combinations of partial order to create total order. \eat{The next phase, called input-parameter value rendering, uses - random, specification constraints, previous response, and previous successful requests for generating value domain for each parameter and performs a t-way covering away principle from combinatorial testing to generate test data for each operation.}}

\section{Conclusion}
\label{sec:conc}

In this paper, we presented a novel black-box API testing algorithm that uses the response message to generate functional test cases. Our objective is to create a reasonable number of test cases that are valid and have high coverage. Central to our technique is the extraction of a variety of constraints from response messages using an intelligent agent. We also augment the specification model with the constraints and generate test cases that satisfy the constraint. Our technique produces the best results when evaluated on 11 benchmark APIs compared to the existing techniques. {\system} generated valid test cases with a very small number of requests due to an intelligent procedure and therefore suitable for cases when API hit is costly. In the future, we would like to combine our greedy procedure with the backtracking procedure for greater benefit. \\

\section{Data Availability}
\label{sec:data}
The detailed results and output dumps obtained for all 4 tools over the 11 benchmark APIs have been made available on repository \footnote{\url{https://anonymous.4open.science/r/ASTRA-B96F}}.
The repository also contains the runnable wheel of {\system} along with the instructions to run it.

\bibliographystyle{ACM-Reference-Format}
\bibliography{bibfile}

\end{document}